\documentclass[aps,pra,reprint,superscriptaddress,nofootinbib]{revtex4-1}
\usepackage{amsmath}
\usepackage{amssymb}
\usepackage{mathrsfs}
\usepackage{graphicx}
\usepackage[caption=false]{subfig}
\usepackage{enumitem}
\usepackage[normalem]{ulem}
\usepackage[percent]{overpic}
\usepackage{bm}
\usepackage{rotating}
\usepackage[usenames, dvipsnames]{color}
\usepackage{hyperref}
\usepackage{cancel}
\usepackage{verbatim}
\usepackage{mathtools}
\usepackage{graphicx}
\usepackage{tensor}
\usepackage{verbatim}
\usepackage{MnSymbol}
\usepackage{booktabs}
\graphicspath{ {images/} }

\usepackage{braket}

\newcommand{\ii}{\mathrm{i}}

\renewcommand{\d}{\mathrm{d}}

\newcommand{\tcr}{\textcolor{red}}

\begin{document}
\title{Thermal Contact: Mischief and Time Scales}
\author{Daniel Grimmer}
\email{dgrimmer@uwaterloo.ca}
\affiliation{Institute for Quantum Computing, University of Waterloo, Waterloo, ON, N2L 3G1, Canada}
\affiliation{Dept. Physics and Astronomy, University of Waterloo, Waterloo, ON, N2L 3G1, Canada}

\author{Robert B. Mann}
\email{rbmann@uwaterloo.ca}
\affiliation{Dept. Physics and Astronomy, University of Waterloo, Waterloo, ON, N2L 3G1, Canada}
\affiliation{Institute for Quantum Computing, University of Waterloo, Waterloo, ON, N2L 3G1, Canada}
\affiliation{Perimeter Institute for Theoretical Physics, Waterloo, ON, N2L 2Y5, Canada}

\author{Eduardo Mart\'{i}n-Mart\'{i}nez}
\email{emartinmartinez@uwaterloo.ca}
\affiliation{Institute for Quantum Computing, University of Waterloo, Waterloo, ON, N2L 3G1, Canada}
\affiliation{Dept. Applied Math., University of Waterloo, Waterloo, ON, N2L 3G1, Canada}
\affiliation{Perimeter Institute for Theoretical Physics, Waterloo, ON, N2L 2Y5, Canada}

\begin{abstract}
We discuss what kind of quantum channels can enable thermalization processes. We show that in order to determine a system's temperature, a thermometer needs to have information about the system's local Hamiltonian and not just its state. We illustrate this showing that any temperature measurement protocol that does not resolve the system's local Hamiltonian (such as, e.g., full state tomography) is susceptible to being fooled into measuring any value for the temperature. We establish necessary conditions for thermal contact for quantum systems. Furthermore, we show that the intuitive idea of thermalization emerging out of quickly interacting with the microconstitutents of a thermal reservoir cannot be correct.

%We discuss what kinds of quantum channels can mediate thermal contact between two quantum systems. Considering the zeroth law of thermodynamics, one can view thermal contact as a way of testing whether two systems are at the same temperature. We show that a wide class of interactions cannot robustly perform this temperature comparison task since they are vulnerable to being sabotaged. We show that to avoid this attack, the interaction between the systems must depend non-trivially on their local Hamiltonians. We go on to show that the intuitive thermalization scenario of a quantum system being rapidly bombarded by the constituents of its environment is vulnerable to this attack and so cannot mediate thermal contact.

%We discuss what kinds of quantum channels can enable thermalization processes. We show that in order to determine a system's temperature, a thermometer needs to dynamically gain information about the system's local Hamiltonian and not just its state. This requires some minimum interaction time. Furthermore, we will argue that the intuitive idea of thermalization by quickly interacting with the microconstitutents of a thermal reservoir cannot be correct.
\end{abstract}

\maketitle

\section{Introduction} 
In life, ``all contact is thermal contact'' is a good rule of thumb. Place any two everyday objects in physical contact with each other and one can confidently predict that they will eventually thermalize to a common temperature -- it seems that no knowledge of how exactly they interact is required. Indeed, it appears this phenomenon is highly robust to the particular details of the scenario considered.

A significant amount of work has been done investigating this robustness, characterizing the emergence of thermalization from quantum dynamics \cite{EquillibriumReview,RevModPhys.83.863,0034-4885-75-12-126001,1612-202X-8-7-001,1367-2630-12-5-055006,QIinT}. Indeed,
use is often made of \textit{Collision Models} to model thermalization \cite{PhysRevLett.115.120403,PhysRev.129.1880,PhysRevA.91.020502,PhysRevA.79.022105,PhysRevA.77.052106} in quantum thermodynamics. Such scenarios consider a quantum system repeatedly interacting with (being bombarded by) the constituents of its environment one at a time. In particular the partial swap interaction described in \cite{Scarani2002} is very common \cite{PhysRevLett.113.100603,PhysRevA.75.052110,PhysRevE.97.022111,1367-2630-16-9-095003,PhysRevA.76.062307}.
In contrast with the intuition highlighted above, these studies show that the details underlying thermal contact are critically important.

Perhaps surprisingly, we will show that in the regime of rapid bombardment, such collisional scenarios cannot constitute thermal contact without fine-tuning. In other words the system is almost never driven to thermal equilibrium with its environment by rapid bombardment. As we will argue, the reason for this non-thermalization in rapid bombardment scenarios is that the system does not have time to sense the temperature of its environment. 

To demonstrate this, we will argue that any dynamics underlying thermal contact must ``know'' both systems' local Hamiltonians. We will then show that without fine-tuning the system must ``learn'' these local Hamiltonians dynamically. Applying these arguments to the rapid bombardment scenario described above we will find the system does not spend enough time with each constituents of its environment to ``learn'' its local Hamiltonian; the process which carries that information is ``too long/complex'' and is highly suppressed. Therefore rapid bombardment cannot mediate thermal contact without fine-tuning. Finally, we will provide some example scenarios, including a comparison with \cite{Scarani2002}.

\section{Thermal Contact and Local Hamiltonians} 
Suppose that we have access to a quantum system in a Gibbs state of unknown inverse temperature, $\beta$, with an unknown local Hamiltonian, $\hat{H}$, that is, \mbox{$\rho_\text{th}=\exp(-\beta \hat{H})/Z$} where \mbox{$Z=\text{Tr}(\exp(-\beta \hat{H}))$}. We can ask the following question: Can we determine its temperature without knowledge of its local Hamiltonian?

An easy answer to this question is: ``Simple! Put the system in thermal contact with a thermometer and read off the temperature''. It appears this task does not require knowledge of the system's local Hamiltonian, just a \textit{thermometer}. Since one simply needs a thermometer to determine the system's temperature, it would seem that doing full state tomography (using arbitrarily many copies of the system) is overkill. However, perhaps surprisingly, state tomography alone does not give us enough information to determine the system's temperature.

To see why,  note that any thermal density matrix \mbox{$\rho_\text{th}=\exp(-\beta \hat{H})/Z$} is invariant under the transformation, 
\begin{align}\label{LambdaDef}
\Lambda:
\beta &\to\lambda \, \beta; \quad
\hat{H} \to \hat{H}/\lambda, 
\end{align}
which rescales both the temperature and energy scale of the system. Imagine that some mischievous agent, Loki himself perhaps, breaks into our lab and replaces our system with a new one which has the same density matrix but nonetheless a different temperature\footnote{Such a transformation can indeed be physically made: for instance, Loki may swap-out a magnetic spin for one with a higher temperature while simultaneously increasing the strength of the magnetic field.}. We cannot notice this swap-out by just characterizing the system's density matrix.

We could however detect Loki's trick if we determine the system's energy scale, which Loki necessarily altered in order to maintain the state's density matrix. This energy scale cannot be determined from the density matrix alone. In fact, as we will see it is the \textit{only} information about the local Hamiltonian not available by characterizing the state, $\rho_\text{th}$. We can determine the eigenbasis, $\{\ket{n}\}$, of $\hat{H}$ by diagonalizing $\rho_\text{th}$ since they commute and therefore share an eigenbasis. Moreover, we can partially determine the eigenvalues, $\{E_n\}$, of $\hat{H}$ using thermal detailed balance,
\begin{align}\label{DetailedBalance}
\frac{\bra{n}\rho_\text{th}\ket{n}}{\bra{m}\rho_\text{th}\ket{m}}=\exp(-\beta(E_n-E_m)).
\end{align}
Specifically, we can determine the relative spacing between the eigenvalues of $\hat{H}$ as, 
\begin{align}\label{DetailedBalance2}
\frac{E_n-E_m}{E_j-E_k}=
\frac{\text{log}(\bra{n}\rho_\text{th}\ket{n})-\text{log}(\bra{m}\rho_\text{th}\ket{m})}
{\text{log}(\bra{j}\rho_\text{th}\ket{j})-\text{log}(\bra{k}\rho_\text{th}\ket{k})}.
\end{align}
This fixes the eigenvalues of $\hat{H}$ up to a constant offset (which can be ignored) and, crucially, an unknown scale factor.

Thus, whatever process we use to measure the system's temperature, it must somehow know (or get to know) the precise value of that scale factor; i.e., it must involve the system's local Hamiltonian. Since (as discussed above) we can measure temperatures via thermal contact with a thermometer, the dynamics which underlie thermal contact must involve the system's local Hamiltonian.

%Furthermore, this necessary energy scale cannot be determined by simply watching the system's free evolution, since the system's state (being thermal) is uneffected by its free evolution. Thus determining this energy scale (and therefore the temperature) of a thermal state requires interacting with the system. As we will see for an interaction to constitute thermal contact the dynamics induced by the interaction (not counting the system's free dynamics) must involve the system's local Hamiltonian.

\section{Thermal Contact and Sabotage}
Two systems (A and B) are in thermal contact if they are allowed to exchange heat and do so until they reach a thermal equilibrium, where no more heat flows between them. The zeroth law of thermodynamics states that this notion of thermal equilibrium is transitive and so can be thought of as an equivalence relation. Temperature is then \textit{defined} as any monotonic labeling of these equilibria with the ordering ``$\text{hot} \, > \, \text{cold}$'' defined by the direction of heat flow.

To illustrate, the textbook example of thermal contact is two systems permitted to exchange energy freely with each other while maximizing their total entropy. By a standard calculation, this process reaches its equilibrium when $\d S_A/\d E_A = \d S_B/\d E_B$. From this we get the usual textbook definition of temperature \mbox{$1/T\coloneqq\d S/\d E$}, where we have taken $k=1$. Note this is the temperature that appears in the exponential of a Gibbs state, \mbox{$\rho_\text{th}=\exp(-\hat{H}/T)/Z$}.

The approach we use in this paper is to take the temperature appearing in the Gibbs state as the canonical temperature defined by a process known to be thermal contact. Using this we can then judge whether any interaction constitutes thermal contact or not by whether its equilibrium condition is compatible with these canonical notions of temperature and thermal contact.

Specifically, suppose our two quantum systems are initially uncorrelated and thermal with inverse temperatures $\beta_\text{A}(0)$ and $\beta_\text{B}(0)$. Imagine that these systems interact with each other for a long time and then are separated. This interaction constitutes \textit{thermal contact} only if for all initial temperatures: 
\begin{itemize}
    \item [1)] The final reduced states of both systems are thermal, with inverse temperatures $\beta_\text{A}(\infty)$ and $\beta_\text{B}(\infty)$.
    \item [2)] These temperatures are the same, \mbox{$    \beta_\text{A}(\infty)=\beta_\text{B}(\infty)$}.
    \item [3)] Compatibility with the zeroth law: The (reduced) systems do not evolve when placed in thermal contact if and only if they are initially at the same temperature. That is, $\rho_\text{A}(t)=\rho_\text{A}(0)$ and $\rho_\text{B}(t)=\rho_\text{B}(0)$ for all $t\geq0$ if and only if $\beta_\text{A}(0)=\beta_\text{B}(0)$. 
\end{itemize} 
%Note that these conditions are  necessary (but not sufficient) for an interaction to be considered thermal contact.

As such in a generic thermal contact scenario between A and B, the systems are initially thermal (with respect to their local Hamiltonians, $\hat{H}_\text{A}$ and $\hat{H}_\text{B}$) and uncorrelated, i.e., $\rho_\text{AB}(0)
=\rho_\text{A}(0)\otimes\rho_\text{B}(0)$, with,
\begin{align}
\rho_\text{A}(0)
=\frac{e^{-\beta_\text{A}(0)\hat{H}_\text{A}}}{Z_\text{A}(0)},\qquad
\rho_\text{B}(0)
=\frac{e^{-\beta_\text{B}(0)\hat{H}_\text{B}}}{Z_\text{B}(0)}.
\end{align}
The most general completely positive trace preserving (CPTP) evolution of this joint system is given by
\begin{align}\label{PhiABDynamics}
\rho_\text{AB}(t)
\coloneqq\Phi_\text{AB}(t)[\rho_\text{A}(0)\otimes\rho_\text{B}(0)],
\end{align}
for some CPTP map $\Phi_\text{AB}(t)$. This dynamical map will in general depend on all the details of the interaction between the two systems. This includes both the systems' local Hamiltonians as well as any other interaction details, which we collect into the label $\mathcal{I}$ as,
\begin{align}\label{PhiABDependence}
\Phi_\text{AB}(t)=\Phi_\text{AB}(t,\hat{H}_A,\hat{H}_B,\mathcal{I}).
\end{align}
For instance, $\mathcal{I}$ could include: an interaction Hamiltonian, $\hat{H}_\text{AB}$, time dependent switching functions, spatial smearing functions, probability distributions for various stochastic elements, the initial temperatures of the system, and other particular details.

\begin{comment}
\tcr{{\bf [Dan: Is all of this red block necessary? The referee complained about how slowly we get to our point]} The systems' reduced states at a time $t$ are given by
\begin{align}
\rho_\text{A}(t)
\!&=\!\Phi_\text{A}(t)[\rho_\text{A}(0)]
\!\coloneqq\!\text{Tr}_\text{B}\big(\Phi_\text{AB}(t)[\rho_\text{A}\!(0)\!\otimes\!\rho_\text{B}\!(0)]\big)\\
\nonumber
\rho_\text{B}(t)
\!&=\!\Phi_\text{B}(t)[\rho_\text{B}(0)]\!\coloneqq\!\text{Tr}_\text{A}\big(\Phi_\text{AB}(t)[\rho_\text{A}\!(0)\!\otimes\!\rho_\text{B}\!(0)]\big).
\end{align}
Note that the reduced dynamical maps for A and B depend on the parameters of the interaction as 
\begin{align}
\Phi_\text{A}(t)
&=\Phi_\text{A}(t,\hat{H}_\text{A},\hat{H}_\text{B},\rho_\text{B}(0),\mathcal{I})\\
\Phi_\text{B}(t)
\nonumber
&=\Phi_\text{B}(t,\hat{H}_\text{A},\hat{H}_\text{B},\rho_\text{A}(0),\mathcal{I}).
\end{align}
The final reduced states are given by,
\begin{align}
\rho_\text{A}\!(\infty)
=\Phi_\text{A}\!(\infty)[\rho_\text{A}\!(0)],\!\quad
\rho_\text{B}\!(\infty)
=\Phi_\text{B}\!(\infty)[\rho_\text{B}\!(0)].
\end{align}}
\end{comment}
Note that the final reduced states depend on the parameters of the interaction as,
\begin{align}
\rho_\text{A}(\infty)
&=f_\text{A}(\hat{H}_\text{A},\hat{H}_\text{B},\rho_\text{A}(0),\rho_\text{B}(0),\mathcal{I})\\
\rho_\text{B}(\infty)
\nonumber
&=f_\text{B}(\hat{H}_\text{A},\hat{H}_\text{B},\rho_\text{A}(0),\rho_\text{B}(0),\mathcal{I}).
\end{align}
If the dynamics described by \eqref{PhiABDynamics} constitutes thermal contact then the final reduced states are both thermal,
\begin{align}\label{ThermStateA}
\rho_\text{A}(\infty)
=\frac{e^{-\beta_\text{A}(\infty)\hat{H}_\text{A}}}{Z_\text{A}(\infty)},\qquad
\rho_\text{B}(\infty)
=\frac{e^{-\beta_\text{B}(\infty)\hat{H}_\text{B}}}{Z_\text{B}(\infty)},
\end{align}
with identical temperatures, $\beta_\text{A}(\infty)=\beta_\text{B}(\infty)$. %Note that, as expected, the final states of A and B (and thus their final temperatures) in general depend on their initial temperatures. Crucially, these temperatures always appear in the density matrices, multiplied by the local Hamiltonians. 

As discussed in the previous section, in order to not be vulnerable to Loki's swap-out trick the dynamics underlying thermal contact must be sensitive to the systems' local Hamiltonians (outside of density matrices). As we will now show, there are two ways for the dynamics to accomplish this:  it may either be fine-tuned such that it ``already knows'' the systems' local Hamiltonians (through some co-dependence of the parameters of the interaction) or it may ``learn'' them dynamically.

To exclude the fine-tuning possibility, we now assume that the parameters of the interaction (i.e., $\beta_\text{A}(0)$, $\beta_\text{B}(0)$, $\hat{H}_\text{A}$, $\hat{H}_\text{B}$, and $\mathcal{I}$) are mutually independent. In particular this means that transformations of the form \eqref{LambdaDef} can be performed on either system without affecting the other parameters of the interaction. Using this independence assumption we will now show that the dynamics must depend on the local Hamiltonians to yield thermal contact.

Suppose that $\Phi_\text{AB}(t)$ does not depend on $\hat{H}_\text{A}$ such that, \mbox{$\Phi_\text{AB}(t)=\Phi_\text{AB}(t,\hat{H}_\text{B},\mathcal{I})$}. From this it follows that the systems' final states depend on $\hat{H}_\text{A}$ only through $\rho_\text{A}(0)$,
\begin{align}\label{RhoARhoBIndependent}
\rho_\text{A}(\infty)
&=f_\text{A}(\hat{H}_\text{B},\rho_\text{A}(0),\rho_\text{B}(0),\mathcal{I}),\\
\rho_\text{B}(\infty)
\nonumber
&=f_\text{B}(\hat{H}_\text{B},\rho_\text{A}(0),\rho_\text{B}(0),\mathcal{I}).
\end{align}
Note that both of these states are then invariant under the transformation,
\begin{align}\label{LambdaATrans}
\Lambda_\text{A}: \ &\beta_\text{A}(0)\to\lambda_\text{A} \, \beta_\text{A}(0);\quad\hat{H}_\text{A}\to \hat{H}_\text{A}/\lambda_\text{A},
\end{align}
since \mbox{$\rho_\text{A}(0)=\exp(-\beta_\text{A}(0)\hat{H}_\text{A})/Z(0)$} is invariant under this transformation and by assumption $\hat{H}_\text{B}$, $\beta_\text{B}(0)$, and $\mathcal{I}$ are each independent of both $\hat{H}_\text{A}$ and $\beta_\text{A}(0)$. Using this invariance and \eqref{ThermStateA} we can determine how the final temperatures of the systems transform under $\Lambda_\text{A}$, 
\begin{align}
\Lambda_\text{A}: \ &\beta_\text{A}(\infty)\to\lambda_A \, \beta_\text{A}(\infty);\quad
\beta_\text{B}(\infty)\to \beta_\text{B}(\infty).
\end{align}
The final temperatures of the systems transform differently and will therefore in general not be the same except for a particular fine-tuned choice of $\lambda_\text{A}$.

From an operational point of view, any interaction like the one described above cannot constitute thermal contact since it is vulnerable to Loki's swap-out trick. For instance, suppose that system B is at a standardized temperature and that we have just ``confirmed'' that system A is at the same temperature by placing it in (what we believe is) thermal contact with B, noting that it doesn't evolve, and by invoking property 3) of thermal contact. Suppose that Loki then swaps system A with system $\text{C}=\Lambda_\text{A}[A]$. If we then perform the same procedure using systems $\text{C}$ and B, we would ``confirm'' they are at the same temperature yielding $\beta_{\text{C}}=\beta_\text{B}=\beta_\text{A}$ even though $\beta_{\text{C}}\neq\beta_\text{A}$. %To make this issue as explicit as possible imagine connecting system A to B and B to C via this alleged thermal contact and then placing systems A and C in the textbook/canonical thermal contact described at the start of this section. In this scenario, one would see heat flow only between A and C, in clear violation of the zeroth law. %Hence, such an interaction cannot constitute thermal contact since using it to compare temperatures produces a violation of the zeroth law of thermodynamics.

Hence we can conclude that for an interaction to constitute thermal contact without fine-tuning the joint dynamical map, $\Phi_\text{AB}(t)$, must depend explicitly on $\hat{H}_\text{A}$. The same argument applies reversing the roles of A and B, such that the joint dynamics must depend on $\hat{H}_\text{B}$ as well.

Of course, one may think this condition is trivially satisfied; physically meaningful maps often depend on the systems' local Hamiltonians. Indeed this is true; for instance, consider master equations of the form,
\begin{align}\label{SampleMasterEqs}
\frac{\d}{\d t}\rho_\text{A}(t)
&=\frac{-\ii}{\hbar}[\hat{H}_\text{A},\rho_\text{A}(t)]+\mathcal{D}_\text{A}[\rho_\text{A}(t)],\\
\nonumber
\frac{\d}{\d t}\rho_\text{B}(t)
&=\frac{-\ii}{\hbar}[\hat{H}_\text{B},\rho_\text{B}(t)]+\mathcal{D}_\text{B}[\rho_\text{B}(t)],
\end{align}
where $\mathcal{D}_{\text{A}}$ and $\mathcal{D}_{\text{B}}$ describe the dynamics induced by the interaction. Clearly this time evolution explicitly depends on the local Hamiltonians. Importantly, however, as we will now discuss, dependence on the local Hamiltonians though such local terms (i.e., $[\hat{H}_\text{A},\cdot]$ and $[\hat{H}_\text{B},\cdot]$) is not enough to yield thermal contact. As we will show the induced dynamics ($\mathcal{D}_\text{A}$ and $\mathcal{D}_\text{B}$) themselves must depend on the local Hamiltonians for interaction to produce thermal contact.

To see this consider the scenario where the initial  temperatures of the two systems are  very close. It is reasonable to expect that they will not evolve much as they equilibrate\footnote{Note this is just a (slight) extension of property 3) of thermal contact.}, staying approximately thermal. We claim that by taking their  initial temperatures to be arbitrarily close the reduced states can be made arbitrarily close to thermal states throughout their evolution. Specifically we claim that the local terms in \eqref{SampleMasterEqs} can be made arbitrarily small throughout the interaction.

%We claim there are physical situations where the local terms in \eqref{SampleMasterEqs} can be made arbitrarily small. Specifically, we claim there are situations where the reduced states of both systems are arbitrarily near to thermal states throughout the interaction. The existence and physicality of such situations seem natural in the sense that by taking the initial temperatures of the two systems to be sufficiently near to each other, then for all times $t\geq0$, it is reasonable to expect that the intermediate reduced states of both systems can by made arbitrarily close to their initial states\footnote{Note this is just a (slight) extension of property 3) of thermal contact.}, which are thermal states. 

Therefore in order for \eqref{SampleMasterEqs} to constitute thermal contact the induced dynamics ($\mathcal{D}_\text{A}$ and $\mathcal{D}_\text{B}$) must themselves depend on the local Hamiltonians; the dynamics must learn the local Hamiltonians dynamically. More generally, the above argument shows that for any dynamics to constitute thermal contact it must depend on the systems' local Hamiltonians \textit{even after} assuming the systems are in (or arbitrarily near to) thermal states throughout the interaction. 

At first glance, this improved condition does not seem much more difficult to satisfy. It is true that in a generic scenario, the induced dynamics will generally depend on both systems' local Hamiltonians. However, as we will soon see, the dynamics generated by rapid bombardment (which we intuitively expect to underlie thermalization) in fact fails to meet this condition and therefore cannot yield thermal contact.

As a final note, we reiterate that these results hold as long as there is no fine-tuning, that is as long as the independence assumption holds. Of course scenarios can be constructed that do not satisfy this assumption. For example, consider two identical magnetic spins (\mbox{$\mu_\text{A}=\mu_\text{B}=\mu$}) polarized by the same magnetic field (\mbox{$B_\text{A}=B_\text{B}=B$}) with, \mbox{$\hat{H}_\text{A}=\mu \,  B \, \hat{\sigma}_{z,A}$}, and \mbox{$\hat{H}_\text{B}=\mu \, B \, \hat{\sigma}_{z,B}$}.  In this case Loki's trick cannot address each system individually. Additionally there are situations (for instance phase transitions) where the local Hamiltonians or coupling strengths are temperature dependent. Nonetheless the scope of applicability of these results is very wide and includes all situations where the local Hamiltonians and coupling strengths are set independently by fundamental considerations.

%\tcr{The class of interaction Hamiltonians, which conserve the sum of the local energies and allow for energy transfer between the systems must violate the independence assumption. If $\hat{H}_\text{SA}$ commutes with $\lambda_\text{S}\hat{H}_\text{S}+\lambda_\text{A}\hat{H}_\text{A}$ it must commute with each of $\hat{H}_\text{S}$ and $\hat{H}_\text{A}$ individually, such that it cannot transfer energy.}

\section{Rapid Bombardment is not Thermal Contact} 
Consider a quantum system, S, interacting with an environment, E, composed of infinitely many identical uncoupled quantum systems, $A_i$, (called ancillas). Suppose that the system interacts unitarily with (is bombarded by) these ancillas one at a time, each for a time $\delta t$. 

This scenario constitutes a \textit{Collision Model} and is commonly used in quantum thermodynamics to model thermalization \cite{PhysRevLett.115.120403,PhysRev.129.1880,PhysRevA.91.020502,PhysRevA.79.022105,PhysRevA.77.052106}. It seems natural to expect that the system will (or at least can) be driven to the temperature of its environment by some bombardment process. Informally, we often think of thermalization as a process where a microscopic constituent of the thermal reservoir interacts with the system for a short time, then flies away and a fresh microconstitutent of the environment interacts with the system again without holding any memory of previous interaction, and the process is repeated until equilibration. We will prove that this intuition is actually wrong, without fine tuning, rapid bombardment cannot mediate thermal contact.

Concretely, suppose that the system and environment are initially uncorrelated and thermal (with respect to their local Hamiltonians, $\hat{H}_\text{S}$ and $\hat{H}_\text{E}$) with inverse temperatures $\beta_\text{S}(0)$ and $\beta_\text{E}(0)$. Since the ancillas that make up the environment are uncoupled, we have that $\hat{H}_\text{E}=\sum_i\hat{H}_{\text{A}_i}$ where $\hat{H}_{\text{A}_i}=\hat{H}_\text{A}$ are the local Hamiltonians of each ancilla. From this it follows that each ancilla is in the state \mbox{$\rho_\text{A}(0)=\exp(-\beta_\text{E}(0)\hat{H}_{\text{A}})/Z_{\text{A}}(0)$} until its interaction with the system.

Therefore each time the system interacts with an ancilla it is updated by the map, 
\begin{align}
\phi(\delta t)[\rho_\text{S}]
=\text{Tr}_\text{A}\big(e^{-\ii \, \delta t \, \hat{H}/\hbar} \ \rho_\text{S}\otimes \rho_\text{A}(0) \ e^{\ii \, \delta t \, \hat{H}/\hbar}\big),
\end{align}
where $\hat{H}
=\hat{H}_\text{S}\otimes\openone_\text{A}
+\openone_\text{S}\otimes\hat{H}_\text{A}
+\hat{H}_\text{SA}$. Thus at a time $t=n \, \delta t$ the state of the system is given by $n$ applications of $\phi(\delta t)$ to the initial state \mbox{$\rho_\text{S}(n \, \delta t)
=\phi(\delta t)^n[\rho_\text{S}(0)]$}.

Using the rapid repeated interaction formalism developed in \cite{Grimmer2016a} and \cite{Grimmer2017a} we can construct an interpolation scheme between the discrete time points $t=n\,\delta t$. Specifically we can construct the unique interpolation scheme which: 1) exactly matches the discrete dynamics, 2) is time-local and time-independent, and 3) converges as $\delta t\to0$. The interpolated dynamics is given by the master equation $\frac{\d}{\d t}\rho_\text{S}(t)
=\mathcal{L}_{\delta t}[\rho_\text{S}(t)]$, where $\mathcal{L}_{\delta t}\coloneqq\frac{1}{\delta t}\text{Log}(\phi(\delta t))$,
called the effective Liouvillian, generates time translations for the system. 

Since we are interested in the rapid bombardment regime\footnote{We note that this rapid bombardment regime is relevant for investigations of Strong Local Passivity \cite{StrongLocalPassivity}.} it is useful for us to expand $\phi(\delta t)$ as a series in $\delta t$, as \mbox{$\phi(\delta t)
=\openone
+\delta t \, \phi_1
+\delta t^2 \, \phi_2
+\delta t^3 \, \phi_3
+\dots$}
where,
\begin{align}\label{phinDef}
&\phi_1[\rho_\text{S}]
=\frac{-\ii}{\hbar} \, 
\text{Tr}_\text{E}\Big(
[\hat{H},\rho_\text{S}\otimes \rho_\text{A}(0)]\Big),\\
\nonumber
&\phi_2[\rho_\text{S}]
=\frac{1}{2!}\Big(\frac{-\ii}{\hbar}\Big)^2
\text{Tr}_\text{E}\Big(
[\hat{H},[\hat{H},\rho_\text{S}\otimes \rho_\text{A}(0)]]\Big),\\
\nonumber
&\phi_3[\rho_\text{S}]
=\frac{1}{3!}\Big(\frac{-\ii}{\hbar}\Big)^3
\text{Tr}_\text{E}\Big(
[\hat{H},[\hat{H},[\hat{H},\rho_\text{S}\otimes \rho_\text{A}(0)]]]\Big),
\end{align}
etc. From this expansion we can expand $\mathcal{L}_{\delta t}$ as a series as \mbox{$
\mathcal{L}_{\delta t}
=\mathcal{L}_0
+\delta t \, \mathcal{L}_1
+\delta t^2 \, \mathcal{L}_2
+\delta t^3 \, \mathcal{L}_3
+\dots$} where,
\begin{align}
\mathcal{L}_0
&=\phi_1,\\
\mathcal{L}_1
\nonumber
&=\phi_2-\frac{1}{2}\phi_1{}^2,\\
\mathcal{L}_2
\nonumber
&=\phi_3
-\frac{1}{2}(\phi_1\phi_2+\phi_2\phi_1)
+\frac{1}{3}\phi_1{}^3,
\end{align}
etc. In \cite{Layden:2015b} and \cite{Grimmer2016a} the first terms, $\mathcal{L}_0$ and $\mathcal{L}_1$, were computed and analyzed in detail. Specifically, it was shown in \cite{Grimmer2016a} that the common technique of taking a continuum limit $\delta t\to0$ along with an diverging interaction strength, $g$, such that $g^2\delta t=\text{const}$ is equivalent to only considering $\mathcal{L}_0+\delta t\,\mathcal{L}_1$ in the above expansion.

More generally, in the rapid bombardment regime (when $\delta t \, E/\hbar\ll1$ where $E$ is the energy scale of $\hat{H}$) it often suffices to study the lowest order terms in this series. For example, if $\mathcal{L}_0$ and $\mathcal{L}_1$ determine a unique fixed point for the dynamics (i.e., $\mathcal{L_0}+\delta t\mathcal{L}_1$ is full rank) then for small enough $\delta t$ all higher order approximations will also have a unique attractive fixed point. Moreover these higher order fixed points are perturbatively near to the lower order ones for small enough $\delta t$. If such a fixed point is established without knowledge of the systems' local Hamiltonians it cannot be the at the temperature of the environment and thus the dynamics cannot be thermal contact.

This raises the question: at what orders in $\delta t$ can $\mathcal{L}_{\delta t}$ constitute thermal contact? As we will see $\mathcal{L}_0$ and $\mathcal{L}_1$ do not depend on the local Hamiltonian of the ancillas (and so cannot constitute thermal contact) whereas $\mathcal{L}_2$  generically   depends on both $\hat{H}_\text{S}$ and $\hat{H}_\text{A}$.

Using the linearity of the partial trace and the commutator, we can see that the $n^{th}$ term in \eqref{phinDef} involves all the ways of picking one of $\hat{H}_\text{S}$, $\hat{H}_\text{A}$, or $\hat{H}_\text{SA}$ for each of the $n$ copies of $\hat{H}$ appearing in the expressions given by \eqref{phinDef}. We now systematically analyze each of these possible combinations which contain $\hat{H}_\text{A}$. 

Using the cyclic property of partial trace, one finds that all terms with $\hat{H}_\text{A}$ in the outermost commutator vanish. Likewise, using the fact that $\rho_\text{A}(0)$ is thermal and therefore commutes with $\hat{H}_\text{A}$ one finds that all terms with $\hat{H}_\text{A}$ in the innermost commutator vanish. Moreover, using the nested commutator identity, \mbox{$[A,[B,C]]=[B,[A,C]]$} if \mbox{$[A,B]=0$}, one can see that all non-vanishing occurrences of $\hat{H}_\text{A}$ must be ``sandwiched'' on either side by an interaction Hamiltonian $H_\text{SA}$ with which it does not commute. Otherwise we could move the  $\hat{H}_\text{A}$ to either end and the term vanishes by the above arguments. 

Thus all terms  depending on $\hat{H}_\text{A}$ in $\phi_1$ and $\phi_2$ vanish. Since $\mathcal{L}_0$ and $\mathcal{L}_1$ are constructed from $\phi_1$ and $\phi_2$ they do not depend on $\hat{H}_\text{A}$ either. Thus if a unique fixed point is established by $\mathcal{L}_0$ and $\mathcal{L}_1$, the dynamics cannot be thermal contact without fine-tuning.

In $\phi_3$ (and therefore in $\mathcal{L}_2$) we find the first term which depends on $\hat{H}_\text{A}$ and doesn't vanish, namely \mbox{$\text{Tr}_\text{A}\Big([\hat{H}_\text{SA},[\hat{H}_\text{A},[\hat{H}_\text{SA},\rho_\text{S}\otimes \rho_\text{A}(0)]]]\Big)$}. We note that $\mathcal{L}_2$ also depends on $\hat{H}_\text{S}$ non-trivially through terms like \mbox{$\text{Tr}_\text{A}\Big([\hat{H}_\text{SA},[\hat{H}_\text{S},[\hat{H}_\text{SA},\rho_\text{S}\otimes \rho_\text{A}(0)]]]\Big)$}.

We can find an explanation for why $\hat{H}_\text{A}$ doesn't show up until $\mathcal{L}_2$ by interpreting $[\hat{H}_X,\cdot]$ as evolution with respect to $\hat{H}_X$. Doing this we can see that the simplest/shortest process carrying information about the ancilla's local Hamiltonian (and therefore its temperature) is to:
\begin{itemize}
\item [1)] Interact with it (so it is not thermal anymore)
\item [2)] Let it evolve freely (bringing in its energy scale)
\item [3)] Interact with it again (to get the information out).
\end{itemize}
In the rapid bombardment regime this process ``takes too long'' and is therefore highly suppressed.

 To make this more concrete, we can consider the following scenario. Suppose that we have many pairs of thermal systems, $A$ and $B$, and we suspect that Loki carries out his swap-out trick on our $B$ systems as
\begin{align}
\Lambda_\text{B}: \ &\beta_\text{B}(0)\to\lambda \, \beta_\text{B}(0);\quad\hat{H}_\text{B}\to \hat{H}_\text{B}/\lambda,
\end{align}
for some $\lambda$. If we couple $N$ of our pairs of systems together, each for a time $\delta t$, and then measure the $A$ systems, how accurate  an (unbiased) estimate can we make about $\lambda$? For any such measurement and data processing procedure, the variance of this estimator is bounded by the Cramer-Rao Theorem to be \cite{QuantumCRBound}
\begin{align}\label{CramerRao}
\text{Var}(\lambda)\geq\frac{1}{N\,F(\lambda,\delta t)} \end{align}
where $F(\lambda,\delta t)$ is the Fisher information about $\lambda$ in each of the $A$ systems after a time $\delta t$. We will now investigate the scaling of the Fisher information about $\lambda$ for small $\delta t$.

 As above let us assume that the systems interact unitarily such that after a time $\delta t$, the state of each system $A$ is
\begin{align}
\nonumber
\phi(\delta t)[\rho_\text{A}(0)]
=\text{Tr}_\text{B}\big(e^{-\ii \, \delta t \, \hat{H}/\hbar} \ \rho_\text{A}(0)\otimes \rho_\text{B}(0) \ e^{\ii \, \delta t \, \hat{H}/\hbar}\big),
\end{align}
where $\hat{H}
=\hat{H}_\text{A}\otimes\openone_\text{B}
+\openone_\text{A}\otimes\hat{H}_\text{B}
+\hat{H}_\text{AB}$. The Fisher information in this state about $\lambda$ is given by 
\begin{align}
F(\lambda,\delta t)=\text{Tr}(L^2\rho_\text{A}(\lambda,\delta t))
\end{align}
where $L$ is the symmetric logarithmic derivative of $\rho_\text{A}(\lambda,\delta t)$ with respect to $\lambda$ defined by,
\begin{align}
\frac{\d}{\d\lambda}\rho_\text{A}(\lambda,\delta t)
=\frac{\rho_\text{A}(\lambda,\delta t)\,L
+L\,\rho_\text{A}(\lambda,\delta t)}{2}.
\end{align}
As noted above, $\phi(\delta t)$ doesn't depend on $\lambda$ until $\phi_3$ such that the left hand side is $O(\delta t^3)$. Since $\rho_\text{A}(\lambda,\delta t)=\rho_\text{A}(0)+O(\delta t)$ we must therefore have $L=O(\delta t^3)$. This in turn implies that $F(\lambda,\delta t)=O(\delta t^6)$. From \eqref{CramerRao} we can thus see that doing more interactions of shorter duration (e.g., $N\to2N$ and $\delta t\to\delta t/2$) results in significantly less information about $\lambda$.

\section{Illustrative Examples}
First let us briefly review the widely used partial swap model of thermalization \cite{Scarani2002}, in which a two level system, S, with \mbox{$\hat{H}_\text{S}
=E \, \hat{\sigma}_\text{S,z}$} interacts with a series of thermal ancillas, A, with \mbox{$\hat{H}_\text{A}
=E \, \hat{\sigma}_\text{A,z}$} and inverse temperature $\beta_\text{A}$. Each interaction has the systems evolve for a duration $\delta t$ under the Hamiltonian \mbox{$\hat{H}_\text{sw}=\hbar \, J(\hat{\openone}_\text{SA}+ \hat{\bm{\sigma}}_\text{S}\cdot\hat{\bm{\sigma}}_\text{A})/2$}. This implements a partial swap unitary, \mbox{$U(\delta t) = \cos(J\,\delta t) \, \openone
-\ii \sin(J\,\delta t) \, U_\text{Sw}$}
where $U_\text{Sw}$ swaps the states of $S$ and $A$ as \mbox{$U_\text{sw}(\ket{S}\otimes\ket{A})=\ket{A}\otimes\ket{S}$}. If $J \, \delta t\neq n \, \pi$, then repeatedly interacting with these ancillas drives the system to the state \mbox{$\rho_\text{S}(\infty)=\rho_\text{A}(0)$}, such that $\beta_\text{S}(\infty)=\beta_\text{A}$. Thus we find thermalization even in the rapid bombardment regime, when $J \, \delta t\ll1$.

But the above setup is fine-tuned. Allow detuning such that $E_\text{S}\neq E_\text{A}$, we still find that \mbox{$\rho_\text{S}(\infty)=\rho_\text{A}(0)$}, but this now means $E_\text{S} \, \beta_\text{S}(\infty)=E_\text{A} \, \beta_\text{A}$. This situation produces thermal contact if and only if $E_\text{S}\neq E_\text{A}$. Any detuning between the systems will cause them not to thermalize.

\begin{comment}
Consider the example in \cite{Grimmer2016a} of a spin qubit, S, rapidly bombarded by an environment of other spin qubits, A, with
\begin{align}
\hat{H}_\text{S}
&=\frac{\hbar\omega_\text{S} \, \hat{\sigma}_\text{S,z}}{2} , \ 
\hat{H}_\text{A}
=\frac{\hbar\omega_\text{A} \, \hat{\sigma}_\text{A,z}}{2} , \, \ 
\hat{H}_\text{SA}
= J \, \hat{\bm{\sigma}}_\text{S}\cdot\hat{\bm{\sigma}}_\text{A} \; .
\end{align}
\tcr{Note that this interaction implements the partial swap interaction found to yield thermalization in \cite{Scarani2002}. This interaction is commonly used in the quantum thermodynamics community to model thermalization \cite{}}. Assuming that the ancillas are thermal, $\mathcal{L}_0$ and $\mathcal{L}_1$ yield a unique fixed point for the system dynamics. Specifically the system is driven to have the same Bloch vector and density matrix as the ancilla qubits, \mbox{$\bm{a}_\text{S}(\infty)=\bm{a}_\text{A}(0)$} and \mbox{$\rho_\text{S}(\infty)=\rho_\text{A}(0)$}. This does not mean that the system is at the same temperature as the ancillas; rather $\hbar\omega_\text{S} \, \beta_\text{S}(\infty)
=\hbar\omega_\text{A} \, \beta_\text{A}$.
In this situation, the system is driven to the   temperature of its environment if and only if $\omega_\text{S}=\omega_\text{A}$, in violation of the independence assumption. \tcr{In \cite{Scarani2002} it was assumed that $\omega_\text{S}=\omega_\text{A}$. Here we see that any detuning between the spin qubits will yield unequal temperatures.}
\end{comment}

Next consider  an harmonic oscillator, S, rapidly bombarded by an environment of other harmonic oscillators, A, with local Hamiltonians, \mbox{$
\hat{H}_\text{S}
=\hbar\omega_\text{S}(\hat{n}_\text{S}+1/2)$} and 
\mbox{$\hat{H}_\text{A}
=\hbar\omega_\text{A}(\hat{n}_\text{A}+1/2)$} and a generic quadratic interaction Hamiltonian,
\begin{align}
\hat{H}_\text{SA}
=\begin{pmatrix}
\hat{x}_\text{S} & \hat{p}_\text{S}
\end{pmatrix}
\begin{pmatrix}
g_{xx} & g_{xp}\\
g_{px} & g_{pp}
\end{pmatrix}
\begin{pmatrix}
\hat{x}_\text{A} \\ \hat{p}_\text{A}
\end{pmatrix}
=\hat{\bm{X}}_\text{S}^\intercal G 
\hat{\bm{X}}_\text{A}\; .
\end{align}
Assuming that the system and ancillas are each initially in thermal states it was shown in  \cite{ArXivGrimmer2017c} that $\mathcal{L}_0$ and $\mathcal{L}_1$ produce a unique fixed point for the system dynamics. As discussed in the previous section, this dynamics therefore cannot produce thermal contact without fine-tuning. 

Specifically, the system is driven to the thermal state with
\begin{align}
\nu_\text{S}(\infty)=\frac{\text{Tr}(G^\intercal G)}{2\, \text{det}(G)}\nu_\text{A}; \quad \nu_X=\frac{\text{exp}(\hbar\omega_X \beta_X)+1}{\text{exp}(\hbar\omega_X \beta_X)-1},
\end{align}
where $\nu_X$ is a temperature monotone. In order for this interaction to constitute thermal contact we must have \mbox{$\beta_S(\infty)=\beta_A$}. Even when S and A are fine-tuned such that $\omega_\text{S}=\omega_\text{A}$ and we only need $\nu_\text{S}(\infty)=\nu_\text{A}$, we only have thermal contact for a very specific family of interactions, those with $\text{Tr}(G^\intercal G)=2\, \text{det}(G)$. The equilibration properties of such couplings are studied in \cite{ArXivGrimmer2017c}. On the other hand, if S and A are detuned (with $\omega_\text{S}\neq\omega_\text{A}$) then the interaction Hamiltonian must depend each systems' energy scales and the ancillas temperature as $G=G(\omega_\text{S},\omega_\text{A},\beta_\text{A})$ in a very specific fine-tuned way to yield thermal contact.

Finally, as a fun sanity check, consider a molecule placed in the air at room temperature ($T=300$ K) interacting with  nitrogen molecules ($m=28$ amu) via a Van der Waals interaction (with energy scale $E=10^{-20}$ J) as it crosses their Van der Waals radius ($r=2.25 A$). We can estimate the duration of each interaction as,
\begin{align}
\delta t
=\frac{2 \, r}{v_\text{rms}}
=\frac{2 \, r}{\sqrt{3 \, kT/m}}
=0.87 \text{ ps},
\end{align}
such that $\delta t \,  E/\hbar=83$. This is not much less than 1. Thus the interactions in the air are (thankfully) long enough to sense the ambient temperature.  The rapid bombardment regime discussed here occurs at Zeno-like time scales wherein every individual interaction the system only varies perturbatively.
%Sun: For $m_H=1$ amu, $T=15\times 10^6 K$, $P=2\times 10^{11}$ atm we have $\delta t \,  E/\hbar=1$ for $E=10.7$ eV. If $E\ll 10.7$ eV the interaction cannot sense the temperature of its environment.

\section{Conclusion} In order to determine a system's temperature, a thermometer (or any process constituting thermal contact) must necessarily gain information about the system's local Hamiltonian and not just its state. Our results lay out a set of requirements for an interaction to constitute thermal contact. In doing so, we have shown that the intuitive idea of thermalization emerging out of a rapid  bombardment of the microconstitutents of a thermal reservoir with a system cannot yield thermalization.

\acknowledgments
The authors would like to thank Valerio Scarani, Stefan Nimmrichter and Gabriel Landi for helpful discussion. EMM and RBM acknowledge support through the Discovery program of the Natural Sciences and Engineering Research Council of Canada (NSERC). E. M-M. is partially funded by his Ontario Early Researcher award. DG acknowledges support by NSERC through the Vanier Scholarship.

\bibliography{references}

\end{document}